\documentstyle[aps,prd,preprint,tighten,floats,epsfig]{revtex}

\newcommand{\EUJ}[1]{ Eur.\ Phys.\ J.\ {\bf #1}}

\def\MSbar {\hbox{$\overline{\hbox{\tiny MS}}$}}
\def\barMS {\overline{\rm MS}}
\def\nf3{}

\begin{document}
\title{On the Behavior of the Effective QCD\\ Coupling
$\alpha_{\tau}(s)$ at Low Scales\thanks{Research partially supported
by the Department of Energy under contract DE--AC03--76SF00515 and the
Swedish Research Council under contract F 620-359/2001}}
\author{Stanley J.~Brodsky, Sven~Menke$^1$, and Carlos Merino$^2$}
\address{Stanford Linear Accelerator Center, Stanford University,
Stanford, California 94309} \address{$^1$ now at Max-Planck-Institut
f{\"u}r Physik, F{\"o}hringer Ring 6, 80805 M{\"u}nchen, Germany}
\address{$^2$ now at Department of Particle Physics, University of
Santiago de Compostela, 15706 Santiago de Compostela, Spain}
\author{Johan Rathsman} \address{High Energy Physics, Uppsala
University, Box 535, S-751 21 Uppsala, Sweden}
\preprint{\vbox{\flushright US-FT/4-02, MPI-PhE/2002-16,
SLAC-PUB-9559, TSL/ISV-2002-0267 \\ December 2002} }

\maketitle

\begin{abstract}
The hadronic decays of the $\tau$ lepton can be used to determine the
effective charge $\alpha_\tau(m^2_{\tau^\prime})$ for a hypothetical
$\tau$-lepton with mass in the range
$0<m_{\tau^\prime}<m_{\tau}$. This definition provides a fundamental
definition of the QCD coupling at low mass scales.  We study the
behavior of $\alpha_{\tau}$ at low mass scales directly from first
principles and without any renormalization-scheme dependence by
looking at the experimental data from the OPAL Collaboration.  The
results are consistent with the freezing of the physical coupling at
mass scales $s = m^2_{\tau^\prime}$ of order $1\,{\rm GeV}^2$ with a
magnitude $\alpha_\tau \sim 0.9 \pm 0.1$.
\end{abstract}

\vfill \centerline{Submitted to Physical Review D} \vfill
\newpage

\section {Introduction}

One of the major uncertainties in making reliable predictions in
quantum chromodynamics (QCD) is to understand the theory at low
momentum scales where the coupling becomes large and non-perturbative
effects become important.  In fact, it is well known that perturbation
theory itself is not well defined in the infrared domain, since the
perturbative series is asymptotic.  For infrared-safe quantities the
non-perturbative effects can be parameterized as power corrections of
the form $c_n(\Lambda/Q)^n$, where $\Lambda$ is the QCD-scale and $Q$
is the hard scale of the process considered.  The coefficients $c_n$
cannot be calculated in general, but in some cases such as the
event-shapes in ${\rm e}^+{\rm e}^-$ annihilation, one can determine
the dominating power $n$ and also find relations between the $c_n$ for
different observables by using the operator product
expansion~\cite{Shifman:bx} or renormalon calculus (for a review see
ref.~\cite{Beneke:1998ui}).

The behavior of fixed-order perturbation theory at low momentum scales
is governed by the running coupling $\alpha_{\rm s}(k^2)$.  The
conventional coupling $\alpha_{\MSbar}(k^2)$, defined using the
modified minimal subtraction renormalization scheme and dimensional
regularization, is analytically singular at a scale $k^2
=\Lambda^2_{\MSbar}$.  This nonphysical behavior leads to a number of
difficulties, including renormalon $n!$ growth of the coefficients
when one makes perturbative expansions of physical observables in the
$\overline{\rm MS}$ scheme.  These problems can be traced to the fact
that integrals over the running coupling which appear in bubble graphs
are ill-defined due to the non-analyticity of the $\barMS$ coupling.

There also exists the possibility to define a running coupling which
stays finite in the infrared. One such example is the ``time-like"
effective coupling which is used in the dispersive
approach~\cite{Beneke:1994qe,Ball:1995ni,Dokshitzer:1995qm}. In this
approach, an observable is written as an integral over a process
dependent characteristic function times a universal ``time-like"
effective coupling. The idea is that such a coupling may give an
effective measure of the interaction at low
scales~\cite{Dokshitzer:1995qm}. Power corrections can then be written
as process dependent moments over the effective coupling.

An alternative procedure is to define the fundamental coupling of QCD
from a given physical
observable~\cite{Grunberg:1980ja,Grunberg:1982fw}. These couplings,
called effective charges, are all-order resummations of perturbation
theory and include all non-perturbative effects.  Since these physical
charges correspond to the complete theory of QCD, it is guaranteed
that they are analytic and non-singular.  For example, it has been
shown that unlike the $\barMS$ coupling, a physical coupling is
analytic across quark flavor
thresholds~\cite{Brodsky:1998mf,Brodsky:1999fr}. Furthermore, we
expect that a physical coupling should stay finite in the infrared
when the momentum scale goes to zero.  In turn, this means that
integrals over the running coupling are well defined for physical
couplings. An additional question is whether the physical couplings
freeze to a constant value in the infrared.

Once such a physical coupling $\alpha_{\rm phys}(k^2)$ is chosen,
other physical quantities can be expressed as expansions in
$\alpha_{\rm phys}$ by eliminating the $\barMS$ coupling which now
becomes only an intermediary~\cite{Brodsky:1994eh}.  In such a
procedure there are in principle no further renormalization scale
($\mu$) or scheme ambiguities.  The physical couplings satisfy the
standard renormalization group equation for its logarithmic
derivative, ${{\rm d}\alpha_{\rm phys}/{\rm d}\ln k^2} =
\hat{\beta}_{\rm phys}[\alpha_{\rm phys}(k^2)]$, where the first two
terms in the perturbative expansion of the Gell-Mann Low function
$\hat{\beta}_{\rm phys}$ are scheme-independent at leading twist
whereas the higher order terms have to be calculated for each
observable separately using perturbation theory.

Quantum field theoretic predictions which relate physical observables
cannot depend on theoretical conventions such as the choice of
renormalization scheme or scale ($\mu$). The most well-known example
is the perturbative ``generalized Crewther
relation"~\cite{Broadhurst:1993ru} in which the leading twist QCD
corrections to the Bjorken sum rule for polarized deep inelastic
scattering at a given lepton momentum transfer $Q^2$ are related
through a geometric series to the QCD corrections to $R_{{\rm e}^+{\rm
e}^-}$ at a corresponding CM energy squared, $s^* = s^*(Q^2)$,
independent of renormalization scheme,
$(1+\alpha_R(s^*)/\pi)(1-\alpha_{g_1}(Q^2)/\pi)=1$~\cite{Brodsky:1995tb}.
The ratio of the scales $s^*/Q^2$ has been computed to NLO in
PQCD. Such leading-twist predictions between observables are called
``commensurate scale relations" and are identical for conformal and
nonconformal theories~\cite{Brodsky:1994eh}. In addition, the
conformal coefficients are free of the renormalon factorial
growth~\cite{Brodsky:2000cr,Rathsman:2001xe,Gardi:2001wg}

For example, in QED, the Gell-Mann Low running coupling $\alpha_{\rm
QED}(k^2) = {\alpha(k_0^2)\over 1- \Pi(k^2,k_0^2)}$, which is formally
defined from the renormalization of the dressed photon propagator, is
a physical coupling since it could be determined from a measurement of
the part of the potential $V(k^2)$ between two infinitely heavy test
charges which is linear in their charges; i.e., $4\pi\alpha_{\rm
QED}(k^2) = - k^2 V_{\rm lin}(k^2)$.  Using the skeleton
expansion~\cite{Bjorken_Drell}, the coefficients in perturbative
expansions of other physical quantities are identical to that in a
theory which is conformal, since all effects of the non-zero
$\hat{\beta}$ function are already summed into the integrals over the
running coupling.  By the mean value theorem, the same is also true
for the standard perturbative expansion if the scale, at which to
evaluate the coupling, is properly
chosen~\cite{Brodsky:1982gc,Hornbostel:2002af}.

It is not as simple to identify a suitable physical coupling to be
used in the case of QCD.  For a skeleton expansion to be possible, the
Abelian part of the coupling should coincide with the Gell-Mann Low
coupling, since QCD becomes an Abelian theory in the analytic limit
$N_{\rm c} \to 0$ at fixed $C_{\rm F} \alpha_{\rm s}$ and fixed
$n_{\rm f}/C_{\rm F}$~\cite{Brodsky:1997jk}.  A possible candidate for
a physical coupling in QCD which fulfills this requirement is the
$\alpha_V(k^2)$ scheme defined from the potential between two heavy
test color charges~\cite{Susskind:pi}.  However, in contrast to the
Abelian case, this definition is problematic since the ``$H-$graphs"
which arise from gluon exchange diagrams with a horizontal gluon rung
connecting the ``first" and ``last" exchanged gluons have an infrared
sensitivity which depends on the details of the test charge
wavefunction~\cite{Appelquist:es}.  Another possible generalization of
the Gell-Man Low coupling to non-Abelian theories is the ``pinch"
scheme~\cite{pinch1,pinch2,Watson:1996fg} which rearranges the
contributions to scattering amplitudes to insure a structure
$\alpha_{\rm pinch}(k^2) = {\alpha_{\rm pinch}(k_0^2)\over 1-
\Pi(k^2,k_0^2)}$ similar to that of QED.  As in QED, expansions in the
pinch scheme have the same structure as those of a conformal theory.
The pinch charge is a promising physical scheme, but at this time the
complexity of higher order calculations in this scheme and its
indirect connection to measurements has prevented its practical
implementation, although recently there has been attempts to make an
all-orders definition of a QCD effective charge~\cite{Binosi:2002vk}.

In this note we will discuss an alternative definition of a physical
coupling for QCD which has a direct relation to high precision
measurements of the hadronic decay channels of the $\tau^- \to
\nu_\tau {\rm h}^-$.  Details on the extraction of $\alpha_{\rm
s}(m_\tau^2)$ from $\tau$ decays can be found in
refs.~\cite{Ball:1995ni,Braaten:hc,Braaten:1988ea,Narison:1988ni,Braaten:1991qm,LeDiberder:1992fr,Neubert:1995gd,Maxwell:1996ig,Girone:1995xb,Barate:1998uf,OPAL} 
(for some recent developments in the extraction of $\alpha_{\rm
s}(m_\tau^2)$ from $\tau$ decays see for
example~\cite{Korner:2000xk,Milton:2000fi,Cvetic:2001sn,Cvetic:2001ws}).

Let $R_{\tau}$ be the ratio of the hadronic decay rate to the leptonic
one.  Then $R_{\tau}\equiv R_{\tau}^0\left[1+{\alpha_\tau \over
\pi}\right]$, where $R_{\tau}^0$ is the zeroth order QCD prediction,
defines the effective charge $\alpha_\tau$.  Throughout this paper we
will concentrate on non-strange decay modes and thus
$R_{\tau}^0=3S_{\rm EW}|V_{\rm ud}|^2$ where $S_{\rm EW}=1.0194$ is an
electroweak correction term~\cite{Marciano:vm} and $|V_{\rm ud}|^2 =
0.9512\pm0.0008$~\cite{Barnett:1996hr} is the relevant CKM matrix
element.  The data for $\tau$ decays is well-understood channel by
channel, thus allowing a precise separation of vector and axial-vector
decay modes which can therefore be studied separately.

The measured invariant mass spectrum for the non-strange hadronic
decay modes can also be used to study hypothetical $\tau$-leptons with
a smaller mass, $m_{\tau^\prime} <
m_\tau$~\cite{Girone:1995xb,Barate:1998uf,OPAL}. In this way the
$\tau$-decay data allows us to study the behavior of the coupling
$\alpha_{\tau}(s)$ in the region $0\le s \le m_\tau^2$ and address the
question whether this physical coupling freezes in the infrared.

\section{Extracting $\alpha_{\tau}(s)$ from data on $\tau$-decays}

The experimental data on the non-strange hadronic $\tau$-decays can be
used to define the hadronic decay rate normalized to the leptonic one
for a hypothetical $\tau$-lepton with mass in the range
$0<m_{\tau^\prime}<m_{\tau}$ in the following
way~\cite{Barate:1998uf}:
\begin{equation}
R_{\tau}^{\rm V/A}(m_{\tau^\prime}^2) \equiv 12 \pi S_{\rm EW}|V_{\rm
ud}|^2\!\!\!\int\limits_0^{m_{\tau^\prime}^2}\!\!\frac{{\rm
d}s}{m_{\tau^\prime}^2}\left(1-\frac{s}{m_{\tau^\prime}^2}\right)^2
\left[ \left(1+2\frac{s}{m_{\tau^\prime}^2}\right){\rm Im} \Pi_{\rm
V/A}^{(1)}(s) +{\rm Im} \Pi_{\rm V/A}^{(0)}(s) \right],
\end{equation}
where $2\pi{\rm Im} \Pi_{\rm V/A}^{(1)}(s)=v/a(s)$ are the spectral
functions for the measured non-strange hadronic final states with
angular momentum $J=1$. The scalar contribution ${\rm Im} \Pi_{\rm
V/A}^{(0)}(s)$ is assumed to vanish for the vector current and is
given by the single pion pole for the axial current.  The above
definition coincides with the standard definition in the case
$m_{\tau^\prime}=m_{\tau}$.  Note, however, that the right-hand side
above uses $m_{\tau^\prime}$ instead of $m_{\tau}$ not only in the
upper integration limit but also in the kinematic prefactors.  This
way the end-point is suppressed in the same way for the hypothetical
$\tau$-lepton as for the real one~\cite{Barate:1998uf}.  The ratio
$R_{\tau}^{\rm V/A}(m_{\tau^\prime}^2)$ is thus defined as if a
hypothetical lepton of mass $m_{\tau^\prime}$ existed.

Since states with non-zero strangeness can be excluded, all of the
hadrons in the final state can be assumed to arise from $\tau^- \to
\nu_\tau\bar{\rm u}{\rm d}$. We can then define effective charges
$\alpha^{\rm V/A}_{\tau}(m_{\tau^\prime}^2)$ for the vector and
axial-vector decay modes as follows:
\begin{equation}
R_{\tau}^{\rm V/A}(m_{\tau^\prime}^2)\equiv \frac{R_{\tau}^0}{2}
\left[1+{\alpha_{\tau}^{\rm V/A}(m_{\tau^\prime}^2) \over \pi}\right].
\label{RtauVAPQCD}
\end{equation}
Notice that the combination $R_{\tau}^{\rm V}-R_{\tau}^{\rm A}$ does
not receive a perturbative QCD contribution at leading twist; thus if
QCD is correct this contribution should be power law suppressed at
high energies.  It is thus natural to identify the complimentary
$R_{\tau}^{\rm V}+R_{\tau}^{\rm A}$ combination which has canonical
perturbative QCD contributions as the preferred QCD effective charge,
$\alpha_{\tau}(m_{\tau^\prime}^2)$, defined by
\begin{equation}
R_{\tau}(m_{\tau^\prime}^2)=R_{\tau}^{\rm V}(m_{\tau^\prime}^2)+
R_{\tau}^{\rm A}(m_{\tau^\prime}^2)\equiv R_{\tau}^0
\left[1+{\alpha_{\tau}(m_{\tau^\prime}^2) \over \pi}\right].
\label{RtauPQCD}
\end{equation}

For completeness, we also recall how one can use experimental data to
measure the decay ratio of hypothetical $\tau$-leptons for masses well
above $m_\tau$~\cite{Brodsky:1998ua}.  Just as in the case of
$\tau$-leptons one can define a local unintegrated effective charge
$\alpha_R(s)$ directly from the annihilation data:
\begin{equation}
R_{{\rm e}^+{\rm e}^-}(s) = {\sigma({\rm e}^+{\rm e}^- \to {\rm
hadrons})\over \sigma({\rm e}^+{\rm e}^- \to \mu^+ \mu^-)} \equiv
R_{{\rm e}^+{\rm e}^-}^0\left[1+\frac{\alpha_R(s)}{\pi}\right],
\label{alphaR}
\end{equation}
where $R_{{\rm e}^+{\rm e}^-}^0$ is the zeroth order QCD prediction.
If we assume isospin invariance, then the decay ratio of a
hypothetical $\tau$-lepton in the vector channel $R_{\tau}^{\rm
V}(m_{\tau^\prime}^2)$ can be written as a spectral integral with
weight $f(x) = \left(1- x\right)^2 \left(1 + 2x\right)$, where $x= {s
/ m_{\tau^\prime}^2}$, of the annihilation cross section ${\rm
e}^+{\rm e}^- \to \gamma^* \to {\rm hadrons}$ in the isospin $I = 1$
channel.  This allows the measurement of
$\alpha^V_{\tau}(m_{\tau^\prime}^2)$ well above the physical mass of
the $\tau$~\cite{Brodsky:1998ua}.  We thus can relate the
$\alpha^V_{\tau}$ and $\alpha_R^{(I=1)}$ effective charges:
\begin{equation}
\alpha^V_{\tau}(m_{\tau^\prime}^2) = {\displaystyle
{\int^{m_{\tau^\prime}^2}_0 \frac{{\rm d}s}{m_{\tau^\prime}^2}\,
f\left(\frac{s}{m_{\tau^\prime}^2}\right)\alpha^{(I=1)}_R(s)} \over
{\displaystyle \int^{m_{\tau^\prime}^2}_0 \frac{{\rm
d}s}{m_{\tau^\prime}^2} \,f\left(\frac{s}{m_{\tau^\prime}^2}\right)}}.
\label{alphaRf}
\end{equation}
The mean value theorem then implies
\begin{equation}
\alpha^V_{\tau}(m_{\tau^\prime}^2)= \alpha^{(I=1)}_R(s^*),\hspace{2cm}
0\leq s^*\leq m_{\tau^\prime}^2,
\label{MVT}
\end{equation}
a form of commensurate scale relation. In the case of three flavors
($n_{\rm f}=3$) the above relation is still valid to
next-to-next-to-leading order even if one does not restrict oneself to
the $I=1$ channel.  To next-to-leading order in $\alpha_R$ the scale
$s^*$ is given by (see for example~\cite{Brodsky:1994eh})
\begin{equation}\label{eq:csrnlo}
s^* = m_{\tau^\prime}^2 \exp\left[-{19 \over 12} -{169 \over
576}\left(11-\frac{2}{3}n_{\rm f}\right)\frac{\alpha_R}{\pi}\right].
\end{equation} 
Before continuing we also note that as an alternative definition of a
hypothetical $\tau$-lepton with mass above $m_\tau$ one could use
${\rm Im} \Pi_{\rm V}^{(1)}(s)$ measured from $\tau$-decays for the
integration region $0<s<m_\tau^2$ and data from $R_{{\rm e}^+{\rm
e}^-}(s)$ in the $I=1$ channel for the remaining integration region
$m_\tau^2<s<m_{\tau^\prime}^2$.

\begin{figure}[htb]
\centering
\epsfig{file=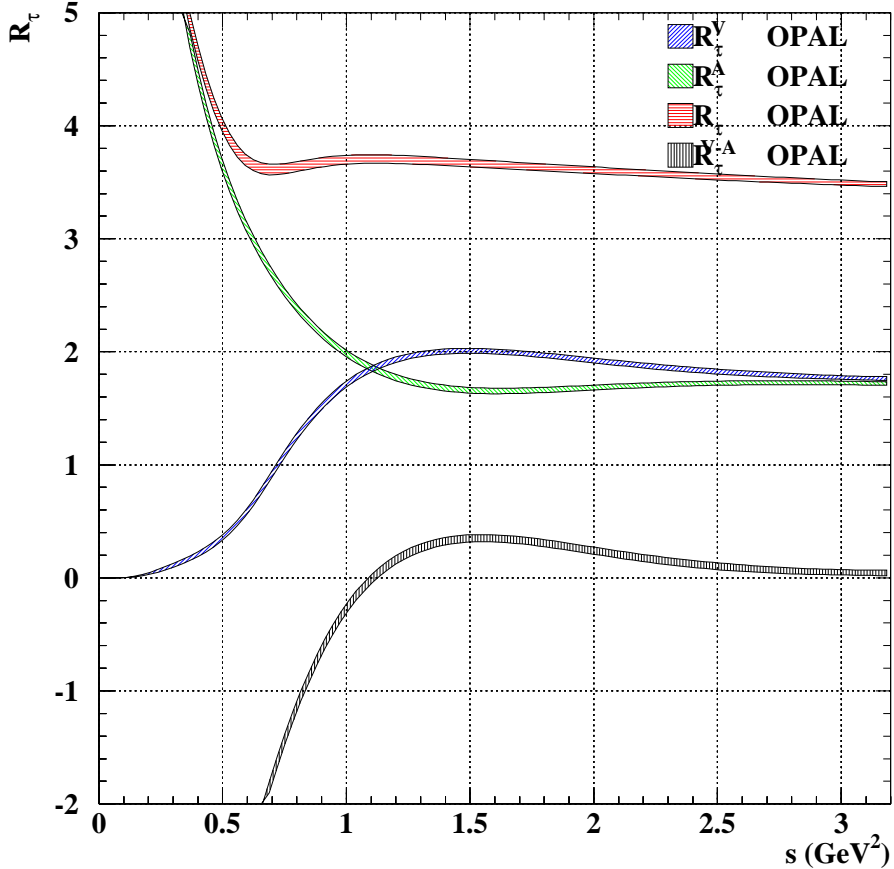,width=0.8\textwidth}
\caption{\em The non-strange hadronic decay rate of a hypothetical
$\mathit{\tau}$ lepton with $\mathit{m_{\tau'}^2 = s}$ versus the
upper integration limit $\mathit{s}$ for the currents $\mathit{V}$,
$\mathit{A}$, $\mathit{V+A}$, and $\mathit{V-A}$.  Error bands include
statistical and systematic errors.}\label{fig:rtau}
\end{figure}

The empirical behavior of the decay ratios $R_{\tau}$, $R_{\tau}^{\rm
V}$, $R_{\tau}^{\rm A}$, and $R_{\tau}^{\rm V}-R_{\tau}^{\rm A}$ using
data on $\tau$-decays as determined by the OPAL collaboration at
LEP~\cite{OPAL,Menke} are shown in fig.~\ref{fig:rtau}.
There are several striking features:
\begin{enumerate}
\item the $R_{\tau}^{\rm V}-R_{\tau}^{\rm A}$ combination tends to
vanish at high scales, showing that only higher dimensional operators
contribute.  This is a highly non-trivial test of
QCD.~\cite{Ioffe:2000ns,Geshkenbein:2001mn}
\item The $R_{\tau}$ contribution has only a slow variation in the low
mass range.
\end{enumerate}

\begin{figure}[htb]
\centering
\epsfig{file=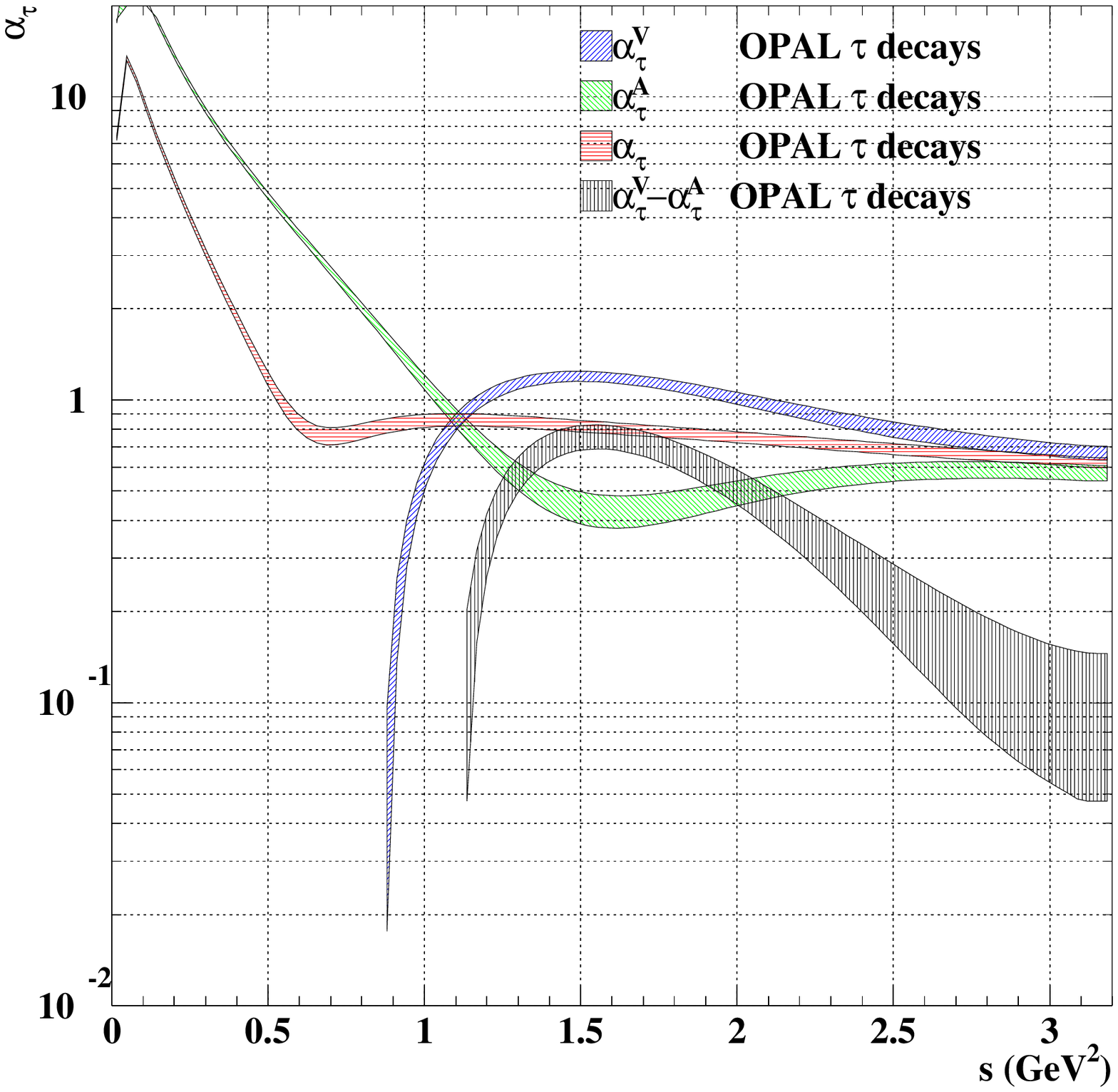,width=0.8\textwidth}
\caption{\em The effective charges for non-strange hadronic decays of
a hypothetical $\mathit{\tau}$ lepton with $\mathit{m_{\tau'}^2 = s}$
versus the upper integration limit $\mathit{s}$ for the currents
$\mathit{V}$, $\mathit{A}$, $\mathit{V+A}$. Also shown is the
difference of the effective charges for the vector and axial-vector
current.  Error bands include statistical and systematic
errors.}\label{fig:alphaeff}
\end{figure}

The corresponding effective charges $\alpha_{\tau}$,
$\alpha_{\tau}^{\rm V}$, and $\alpha_{\tau}^{\rm A}$ as well as the
difference $\alpha_{\tau}^{\rm V}-\alpha_{\tau}^{\rm A}$ are shown in
Fig.~\ref{fig:alphaeff}.

Based on the analysis by the OPAL collaboration~\cite{OPAL}, the
experimental value of the coupling $\alpha_{\tau}(s)=0.621\pm0.008$ at
$s = m^2_\tau$ corresponds to a value of $\alpha_{\MSbar}(M^2_Z) =
(0.117$-$0.122) \pm 0.002$, where the range corresponds to three
different perturbative methods used in analyzing the data.  This
result is, at least for the fixed order and renormalon resummation
methods, in good agreement with the world average
$\alpha_{\MSbar}(M^2_Z) = 0.117 \pm 0.002$~\cite{PDG}. However, from
the figure we also see that the effective charge only reaches
$\alpha_{\tau}(s) \sim 0.9 \pm 0.1$ at $s=1\,{\rm GeV}^2$, and it even
stays within the same range down to $s\sim0.5\,{\rm GeV}^2$.  This
result is in good agreement with the estimate of Mattingly and
Stevenson~\cite{Mattingly:ej} for the effective coupling $\alpha_R(s)
\sim 0.85 $ for $\sqrt s < 0.3\,{\rm GeV}$ determined from ${\rm
e}^+{\rm e}^-$ annihilation, especially if one takes into account the
perturbative commensurate scale relation,
$\alpha_{\tau}(m_{\tau^\prime}^2)= \alpha_R(s^*)$ where, for
$\alpha_R=0.85$, we have $s^* \simeq 0.10\,m_{\tau^\prime}^2 $
according to Eq.~(\ref{eq:csrnlo}).  As we will show in more detail in
the next section, this behavior is not consistent with the coupling
having a Landau pole but rather shows that the physical coupling is
much more constant at low scales, suggesting that physical QCD
couplings are effectively constant or ``frozen" at low scales.

At the same time, it should be recognized that the behavior of
$\alpha_{\tau}(s)$ in the region $s<1\,{\rm GeV}^2$ is more and more
influenced by non-perturbative effects as the scale is lowered.  Even
though the dominant non-perturbative effects cancel in the sum of the
vector and axial-vector contributions as can be seen by looking at the
corresponding effective charges individually.  Looking at
$\alpha_{\tau}^{\rm V}(s)$, we see that it more or less vanishes as
the integration region moves to the left of the two-pion peak in the
hadronic spectrum.  In the same way the behavior of
$\alpha_{\tau}^{\rm A}(s)$ at small scales is governed by the single
pion pole.

\section {Analysis of the infrared behavior of $\alpha_{\tau}(s)$}

In order to be able to analyze the infrared behavior of the effective
coupling $\alpha_{\tau}(s)$ in more detail, we will compare with (a)
the fixed-order perturbative evolution of the $\alpha_{\tau}(s)$
coupling on the one hand, and (b) with the evolution of couplings that
have non-perturbative or all-order resummations included in their
definition.  For the latter case, many different schemes have been
suggested, and we will concentrate on two of them: the one-loop
``time-like" effective coupling $\alpha_{\rm
eff}(s)$~\cite{Beneke:1994qe,Ball:1995ni,Dokshitzer:1995qm}, and the
modified $\tilde{\alpha}_{V}$ coupling calculated from the static
quark potential using perturbative gluon condensate
dynamics~\cite{Hoyer:2000ca}.

The perturbative couplings evolve according to the standard evolution
equation
\begin{equation}\label{eq:foptevol}
\frac{{\rm d}a_{\rm s}(s)}{{\rm d}\ln s} = -\beta_0 a_{\rm s}^2(s)
-\beta_1 a_{\rm s}^3(s) -\beta_2 a_{\rm s}^4(s) -\beta_3 a_{\rm
s}^5(s) - \dots,
\end{equation}
where $a_{\rm s}(s) = \alpha_{\rm s}(s)/(4\pi)$.  The first two terms
in the $\beta$-function, $\beta_0$ and $\beta_1$, are universal at
leading twist whereas the higher order terms are scheme dependent.
Currently the $\beta$-function is known to four loops ($\beta_3$) in
the $\barMS$ scheme and to three loops ($\beta_2$) in the
$\alpha_{\tau}$ scheme.  In the latter case there also exists an
estimate of the four-loop term.  For completeness these terms are
summarized in the appendix.

Fig.~\ref{fig:fopt_comp} shows a comparison of the experimentally
determined effective charge $\alpha_{\tau}(s)$ with solutions to the
evolution equation (\ref{eq:foptevol}) for $\alpha_{\tau}$ at two-,
three-, and four-loop order normalized at $m_\tau$.  It is clear from
the figure that the data on $\alpha_{\tau}(s)$ does not have the same
behavior as the solution of the (universal) two-loop equation which is
singular\footnote{The same divergent behavior would also be seen at
three- and four-loop order in the $\barMS$ scheme where both $\beta_2$
and $\beta_3$ are positive for $n_{\rm f}=3$.} at the scale
$s\simeq1\,{\rm GeV}^2$. However, at three loops the behavior of the
perturbative solution drastically changes, and instead of diverging,
it freezes to a value $\alpha_{\tau}\simeq 2$ in the infrared.  The
reason for this fundamental change is, of course, the negative sign of
$\beta_{\tau,2}$.  At the same time, it must be kept in mind that this
result is not perturbatively stable since the evolution of the
coupling is governed by the highest order term.  This is illustrated
by the widely different results obtained for three different values of
the unknown four loop term $\beta_{\tau,3}$ which are also
shown\footnote{The values of $\beta_{\tau,3}$ used are obtained from
the estimate of the four loop term in the perturbative series of
$R_\tau$, $K_4^{\overline{\rm MS}} =
25\pm50$~\cite{LeDiberder:1992fr}.}.  Still, it may be more than a
mere coincidence that the three-loop solution freezes in the
infrared. Recently it has been argued that $\alpha_R(s)$ freezes
perturbatively to all orders~\cite{Howe:2002rb}.  Given the
commensurate scale relation~(\ref{MVT}) this should also be true
perturbatively for $\alpha_{\tau}(s)$.  It is also interesting to note
that the central four-loop solution is in good agreement with the data
all the way down to $s\simeq1\,{\rm GeV}^2$.

\begin{figure}[htb]
\centering
\epsfig{file=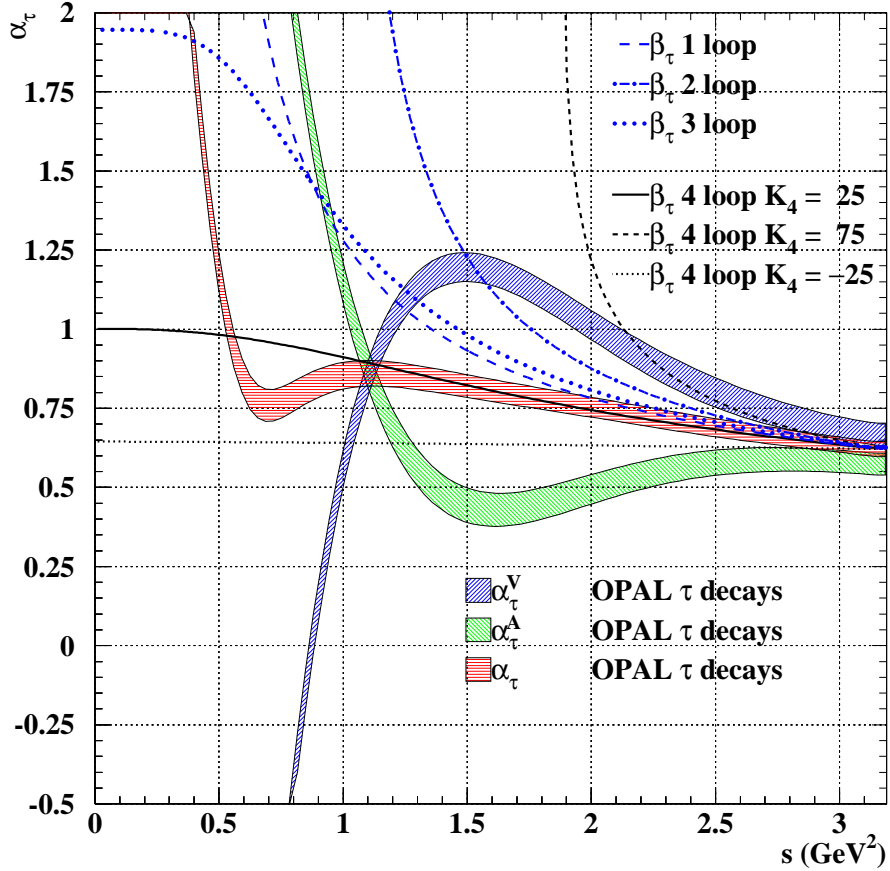,width=0.8\textwidth}
\caption{\em The effective charge $\alpha_{\tau}$ for non-strange
hadronic decays of a hypothetical $\mathit{\tau}$ lepton with
$\mathit{m_{\tau'}^2 = s}$ compared to solutions of the fixed order
evolution equation (\protect\ref{eq:foptevol}) for $\alpha_{\tau}$ at
two-, three-, and four-loop order.  Error bands include statistical
and systematic errors.}\label{fig:fopt_comp}
\end{figure}

The one-loop ``time-like" effective
coupling~\cite{Beneke:1994qe,Ball:1995ni,Dokshitzer:1995qm}
\begin{equation}\label{eq:alphaeff}
\alpha_{\rm eff}(s)=\frac{4\pi}{\beta_0} \left\{\frac{1}{2} -
\frac{1}{\pi}\arctan\left[\frac{1}{\pi}\ln\frac{s}{\Lambda^2}\right]\right\}
\end{equation}
is obtained from the analytic continuation of the one-loop coupling
$\alpha_{\rm s}(Q^2)=(4\pi)/[\beta_0\ln(Q^2/\Lambda^2)]$ which defines
the spectral density ${\rm d}\alpha_{\rm eff}(s)/{\rm d}\ln s = [
\alpha_{\rm s}(-s+i\epsilon) - \alpha_{\rm s}(-s-i\epsilon) ]/ (2 \pi
i)$ in the same way as $R_{{\rm e}^+{\rm e}^-}(s)$ is related to the
Adler $D$-function. The resulting effective coupling is finite in the
infrared and freezes to the value $\alpha_{\rm
eff}(s)={4\pi}/{\beta_0}$ as $s\to 0$.  It is also instructive to
expand the ``time-like" effective coupling for large $s$,
\begin{eqnarray*}
\alpha_{\rm eff}(s) &=&\frac{4\pi}{\beta_0\ln\left(s/\Lambda^2\right)}
\left\{1 -\frac{1}{3}\frac{\pi^2}{\ln^2\left(s/\Lambda^2\right)}
+\frac{1}{5}\frac{\pi^4}{\ln^4\left(s/\Lambda^2\right)} +\ldots
\right\}\\ &=&\alpha_{\rm s}(s)\left\{1
-\frac{\pi^2\beta_0^2}{3}\left(\frac{\alpha_{\rm s}(s)}{4\pi}\right)^2
+\frac{\pi^4\beta_0^4}{5}\left(\frac{\alpha_{\rm s}(s)}{4\pi}\right)^4
+\ldots \right\}.
\end{eqnarray*}
This shows that the ``time-like" effective coupling is a resummation
of $(\pi^2\beta_0^2\alpha_{\rm s}^2)^n$-corrections to the
``space-like" coupling which occurs in the analytic continuation.  The
``time-like" effective coupling can also be defined to higher orders
(for a recent review see~\cite{Shirkov:2001sm}) but the infrared
behavior persists also in these cases.

The form for the finite coupling $\alpha_{\rm eff}$ given in
Eq.~(\ref{eq:alphaeff}) obeys standard PQCD evolution at LO. Thus one
can have a solution for the perturbative running of the QCD coupling
which obeys asymptotic freedom but does not have a Landau singularity.

The evolution of the modified $\tilde{\alpha}_{V}(Q^2,\Lambda_{\rm
gc}^2)$ coupling is to leading order governed by the evolution
equation,
\begin{equation}\label{eq:modvevol}
\frac{{\rm d}\tilde{a}_{V}(Q^2,\Lambda_{\rm gc}^2)}{{\rm d}\ln Q^2} =
-\tilde{\beta}_0(\Lambda_{\rm gc}/Q) \tilde{a}_{V}^2(Q^2,\Lambda_{\rm
gc}^2),
\end{equation}
where
\begin{equation}
\tilde{\beta}_0(\Lambda_{\rm gc}/Q)=5-12\frac{\Lambda_{\rm
gc}^2}{Q^2}+ \left( 12\frac{\Lambda_{\rm gc}^3}{Q^3}
-6\frac{\Lambda_{\rm gc}}{Q} +\frac{3}{2}\frac{Q}{\Lambda_{\rm gc}}
\right) \ln\frac{Q+2\Lambda_{\rm gc}}{Q-2\Lambda_{\rm gc}}
-\frac{2}{3} n_{\rm f}.
\end{equation}
The difference compared to the ordinary evolution equation is thus
that the gluonic contribution to $\beta_0$ freezes out at a scale
$Q\sim\Lambda_{\rm gc}$ -- the scale of the gluon condensate.

\begin{figure}[htb]
\centering
\epsfig{file=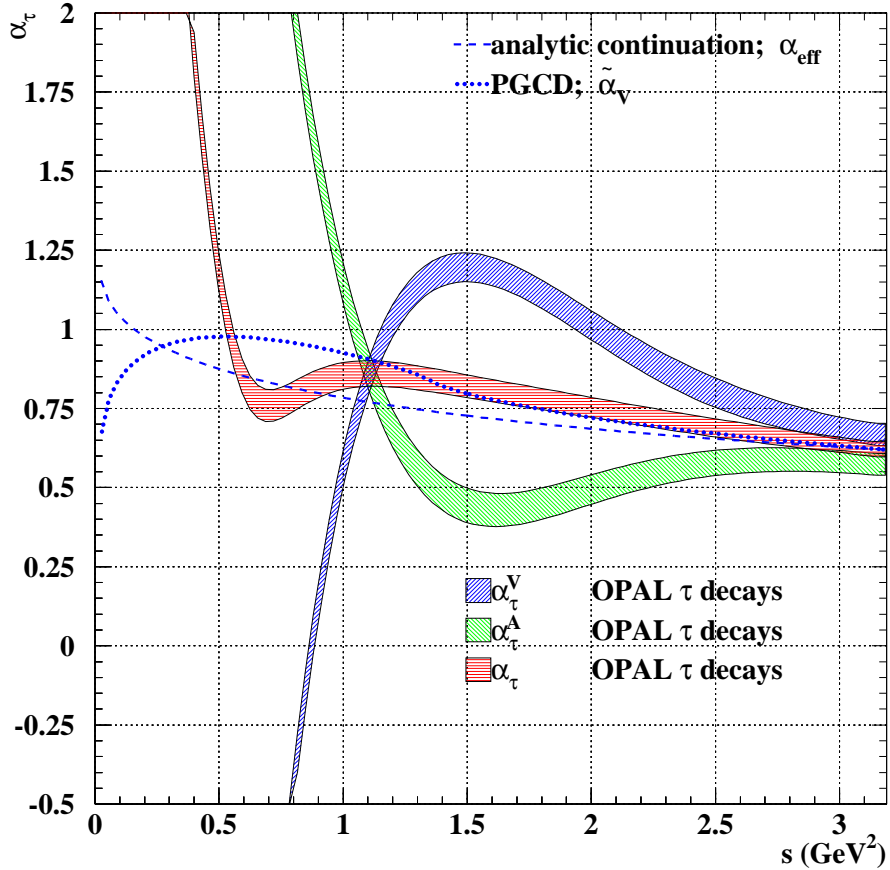,width=0.8\textwidth}
\caption{\em The effective charge $\alpha_{\tau}$ for non-strange
hadronic decays of a hypothetical $\mathit{\tau}$ lepton with
$\mathit{m_{\tau'}^2 = s}$ compared to the one-loop ``time-like"
effective coupling $\alpha_{\rm eff}(s)$ and the modified
$\tilde{\alpha}_{V}$ coupling calculated from the static quark
potential using perturbative gluon condensate dynamics.  Error bands
include statistical and systematic errors.}\label{fig:eff_comp}
\end{figure}

Fig.~\ref{fig:eff_comp} shows a comparison of the experimentally
determined effective charge $\alpha_{\tau}(s)$ with the one-loop
``time-like" effective coupling $\alpha_{\rm eff}(s)$ and the modified
$\tilde{\alpha}_{V}$ coupling calculated from the static quark
potential using perturbative gluon condensate dynamics.  In both cases
the solutions have been normalized to the data at $m_\tau$.  In
addition, when solving the evolution equation for $\tilde{\alpha}_{V}$
(\ref{eq:modvevol}) we have used the value $\Lambda_{\rm gc}=581\,{\rm
MeV}$ which makes the solution to the one-loop evolution equation
(\ref{eq:foptevol}) agree with the data at $m_\tau$ if one sets
$\Lambda=\Lambda_{\rm gc}$.  As can be seen from the figure, the data
on $\alpha_{\tau}(s)$ agrees well qualitatively with both of these
simple examples of freezing couplings down to the scale
$s\simeq1\,{\rm GeV}^2$. Below this scale the non-perturbative effects
in $\alpha_{\tau}(s)$ associated with the single pion pole in the
axial vector current and the double pion peak in the vector current
starts to dominate and makes a direct comparison with models which do
not contain the spectrum of hadrons less meaningful.

\section{Conclusions}

The results of this paper show the advantages of defining the QCD
coupling directly from a physical observable.  The resulting effective
coupling can be defined even at low scales, it is finite and analytic,
and it has no scheme or renormalization scale ambiguities.  As we have
shown, the hadronic decays of the $\tau$ lepton can be used to
determine the effective charge $\alpha_\tau(m^2_{\tau^\prime})$ for a
hypothetical $\tau$-lepton with mass in the range
$0<m_{\tau^\prime}<m_{\tau}$.  The high precision of $\tau$ hadronic
decay data thus provides a precise standard definition for the
fundamental QCD coupling.  QCD predictions for other observables can
then be expressed as functions of this coupling $\alpha_\tau(s)$, thus
relating observable to observable.

An important feature of $\alpha_\tau(s)$ is its apparent near constant
behavior at low mass scales.  The empirical results using the OPAL
data are consistent with the freezing of the physical coupling at mass
scales $s = m^2_{\tau^\prime}$ of order $1\,{\rm GeV}^2$ with a
magnitude $ \alpha_\tau \sim 0.9 \pm 0.1$. These results contrast
dramatically with standard expectations of a divergent coupling based
on the universal two-loop coupling which becomes infinite at small
mass scales. At higher orders of perturbation theory for the beta
function, the behavior of the coupling is scheme dependent and in
addition it is dominated by the highest order term. In the physical
$\alpha_\tau$-scheme the behavior of the coupling in the infrared is
not so clear. At 3-loops the $\alpha_\tau(s)$ coupling has an infrared
fixed point whereas at higher orders the behavior of the coupling is
not known. Estimates of the 4-loop term indicate both an infrared
fixed point as well as a divergent behavior.

Recently it has been argued by Howe and Maxwell~\cite{Howe:2002rb}
that the effective charge $\alpha_R(s)$ has an infrared fixed point to
all orders in perturbation theory.  Given the commensurate scale
relation between $\alpha_\tau$ and $\alpha_R$ this should then also be
true for $\alpha_\tau(s)$. A simple example of the mechanism behind
such a perturbative infrared fixed point is given by the effective
``time-like" one-loop coupling $\alpha_{\rm eff}(s)$ which agrees well
qualitatively with the empirically determined coupling.  Another
mechanism which also gives an infrared fixed point in qualitative
agreement with the data is given by the perturbative gluon condensate
dynamics.  These simple examples show that indeed the empirical
behavior $\alpha_\tau(s)$ is consistent with an infrared fixed point.

As we have discussed in the introduction, effective charges defined
from different observables are related in perturbative QCD to each
other at leading twist through commensurate scale
relations~\cite{Brodsky:1994eh}.  The perturbative coefficients which
appear in these relations are identical to those in conformal QCD;
i.e. the theory defined with a zero $\beta$ function.  The effects of
the nonzero $\beta$ function of physical QCD are absorbed into the
scale of the effective couplings.  Examples are the relations between
$\alpha_R$ and $\alpha_\tau$ and the generalized Crewther relation,
which relates the effective charge defined from the Bjorken sum rule
for $g_1(x,Q^2)$ to the Adler coupling defined from ${\rm e}^+{\rm
e}^-$ annihilation through a geometric series~\cite{Brodsky:1995tb}.

Since $\alpha_\tau(m^2_\tau)$ has effectively constant behavior at
small mass scales, the commensurate scale relations between
$\alpha_\tau$ and other physical couplings imply that at leading twist
all effective charges have a similar fixed-point behavior in the
infrared regime. In particular, our results are consistent with the
infrared freezing of $\alpha_R$ at low $s$ obtained by Mattingly and
Stevenson~\cite{Mattingly:ej}.

The commensurate scale relations~\cite{Brodsky:1994eh} can of course
also be used at larger scales in order to compare experimental results
from different observables. For example, the $\alpha_\tau$ effective
charge can be evolved to the commensurate scale of another observable
which one wants to compare with.  This way the results can be compared
directly, free of theoretical conventions, instead of by first
extracting $\alpha_{\MSbar}(M^2_Z)$ and then comparing.  Beyond
leading twist one can also include power-corrections in the
commensurate scale relations and determine non-perturbative parameters
in a standard way.  Ultimately one could also envision to use a
physical effective charge in the dispersive approach. As we have
shown, the four-loop evolution of $\alpha_\tau$ with
$K_4^{\overline{\rm MS}}=25$ gives a good description of the data down
to $s\sim1\,{\rm GeV}^2$ and could thus be used as an effective model
for the $\alpha_\tau$ coupling in the low-energy regime.

The near constancy of the effective QCD coupling at small scales helps
explain the empirical success of dimensional counting rules for the
power law fall-off of form factors and fixed angle scaling (for a
review see ref.~\cite{Brodsky:1989pv}). As shown in
ref.~\cite{Brodsky:1997dh}, one can calculate the hard scattering
amplitude $T_H$ for such processes~\cite{Lepage:1980fj} without scale
ambiguity in terms of the effective charge $\alpha_\tau$ or $\alpha_R$
using commensurate scale relations. The effective coupling is
evaluated in the regime where the coupling is approximately constant,
in contrast to the rapidly varying behavior from powers of
$\alpha_{\rm s}$ predicted by perturbation theory (the universal
two-loop coupling).  For example, the nucleon form factors are
proportional at leading order to two powers of $\alpha_{\rm s}$
evaluated at low scales in addition to two powers of $1/q^2$; The pion
photoproduction amplitude at fixed angles is proportional at leading
order to three powers of the QCD coupling.  The essential variation
from leading-twist counting-rule behavior then only arises from the
anomalous dimensions of the hadron distribution amplitudes.

The large magnitude that we find $\alpha_\tau \sim 0.9 \pm 0.1$ also
implies a substantially larger normalization for the pion form factor
and other exclusive observables than estimates based on a canonical
value $\alpha_{\rm s} \sim 0.3$. The corresponding larger value of
$\alpha^n_{\rm s}$ associated with the exchange of $n$ gluons in the
hard scattering amplitude could eliminate much of the discrepancy
between data and previous estimates of the PQCD predictions for the
normalization of exclusive hard scattering amplitudes.

\appendix

\section{The $\beta$-function in the $\barMS$
and $\alpha_{\tau}$ schemes}

The perturbative expansion of the $\beta$ function is given by,
\begin{equation}
\frac{{\rm d}a_{\rm s}(s)}{{\rm d}\ln s} = -\beta_0 a_{\rm s}^2(s)
-\beta_1 a_{\rm s}^3(s) -\beta_2 a_{\rm s}^4(s) -\beta_3 a_{\rm
s}^5(s) - \dots,
\end{equation}
where $a_{\rm s}(s) = \alpha_{\rm s}(s)/(4\pi)$.  The first two terms
in the
$\beta$-function~\cite{Gross:1973id,Politzer:fx,Caswell:1974gg,Jones:mm,Egorian:1978zx},
\begin{eqnarray*}
  \beta_0 & = & 11 - \frac{2}{3} n_{\rm f}, \\ \beta_1 & = & 102 -
  \frac{38}{3} n_{\rm f},
\end{eqnarray*}
 are universal at leading twist whereas the higher order terms are
scheme dependent.  In the $\barMS$ scheme the first two scheme
dependent coefficients are
known~\cite{Tarasov:au,Larin:tp,vanRitbergen:1997va}:
\begin{eqnarray}
  \beta_2^{\overline{\rm MS}} & = & \frac{2857}{2} - \frac{5033}{18}
  n_{\rm f} + \frac{325}{54} n_{\rm f}^2, \nonumber\\
  \beta_3^{\overline{\rm MS}} & = & \frac{149753}{6} + 3564\,\zeta_3 -
  \left(\frac{1078361}{162} + \frac{6508}{27} \zeta_3\right) n_{\rm f}
  \nonumber\\ & & + \left(\frac{50065}{162} + \frac{6472}{81}\zeta_3
  \right) n_{\rm f}^2 + \frac{1093}{729} n_{\rm
  f}^3. \label{eq:beta_2}
\end{eqnarray}
In case of the $\alpha_{\tau}$ scheme, $\beta_{\tau,2}$ is known
exactly whereas for $\beta_{\tau,3}$ there are only estimates.

Since the three- and four-loop coefficients in the $\beta$-function
are known in the $\barMS$-scheme, the corresponding coefficients (or
estimates thereof) can be obtained from the perturbative expansion of
$\alpha_{\tau}$ in the $\barMS$-scheme.  Starting from the
perturbative expansion of the Adler D-function for the photon vacuum
polarization:
\begin{equation}
  D(-s) = -4\pi^2s\frac{{\rm d}\Pi^\gamma(-s)}{{\rm d}s} = \sum_f
  Q_f^2\left(1+\sum_{n=1}^4 D_n\frac{\alpha_{\rm
  s}^n(s)}{\pi^n}\right) \label{eq:Adler},
\end{equation}
with the known
coefficients~\cite{Chetyrkin:bj,Dine:1979qh,Celmaster:xr,Gorishnii:1990vf,Surguladze:1990tg}

\begin{eqnarray}
  D_1 & = & 1,\nonumber\\ D_2 & = & \frac{365}{24} - 11\,\zeta_3 -
  \left(\frac{11}{12}-\frac{2}{3}\zeta_3\right)n_{\rm f}, \nonumber\\
  D_3^{\overline{\rm MS}} & = & \frac{87029}{288} -
  \frac{1103}{4}\zeta_3 + \frac{275}{6}\zeta_5 
+\left(-\frac{7847}{216} + \frac{262}{9}\zeta_3 -
    \frac{25}{9}\zeta_5\right)n_{\rm f}\nonumber\\ & &
    +\left(\frac{151}{162} - \frac{19}{27}\zeta_3\right)n_{\rm
    f}^2
+ \left(\frac{55}{72} -
    \frac{5}{3}\zeta_3\right)\frac{\displaystyle\left(
    \sum_fQ_f\right)^2}{\displaystyle N_{\rm c}\sum_fQ_f^2}
\label{eq:dns}
\end{eqnarray}
one can calculate the following expression for $\alpha_\tau$ where
$n_{\rm f}=3$ is assumed for the $\beta_n$:
\begin{eqnarray}
\frac{\alpha_\tau(s)}{\pi} & = & \frac{\alpha_{\MSbar}(s)}{\pi} +
\left(\frac{19}{48}\beta_0{\nf3}+K_2\right)
\frac{\alpha_{\MSbar}^2(s)}{\pi^2}\nonumber\\ & & +
\left\{\left[\frac{265}{1152}-\frac{1}{48}\pi^2\right]
\beta_0^2{\nf3}+\frac{19}{192}\beta_1{\nf3} 
+\frac{19}{24}\beta_0{\nf3}\,K_2 + K_3^{\overline{\rm MS}}\right\}
  \frac{\alpha_{\MSbar}^3(s)}{\pi^3} \nonumber\\ & & +
  \left\{\left[\frac{3355}{18432}-\frac{19}{768}\pi^2\right]
  \beta_0^3{\nf3} 
+ \left[\frac{1325}{9216}-\frac{5}{384}\pi^2\right]
    \beta_0{\nf3}\,\beta_1{\nf3} +
    \frac{19}{768}\beta_2^{\overline{\rm MS}}{\nf3}\right.\nonumber\\
    & & \left.\quad\!\!+
    \left[\left(\frac{265}{384}-\frac{1}{16}\pi^2\right)
    \beta_0^2{\nf3} +
    \frac{19}{96}\beta_1{\nf3}\right]\!K_2
+ \frac{19}{16}\beta_0{\nf3}\,K_3^{\overline{\rm MS}} +
K_4^{\overline{\rm MS}}\right\}\!
\frac{\alpha_{\MSbar}^4(s)}{\pi^4},\label{eq:alphatau}
\end{eqnarray}
where, as is customary in the $\tau$ literature, we use the following
notation for the $D_n$ with ${\rm f}=\{\rm u,d,s\}$:
\begin{eqnarray}
  K_1 & = & D_1 = 1,\nonumber\\ K_2 & = & D_2(n_{\rm f}=3) =
  \frac{299}{24} - 9\,\zeta_3, \nonumber\\ K_3^{\overline{\rm MS}} & =
  & D_3^{\overline{\rm MS}}(n_{\rm f}=3) = \frac{58057}{288} -
  \frac{779}{4}\zeta_3 + \frac{75}{2}\zeta_5 \nonumber\\
  K_4^{\overline{\rm MS}} & \sim & 25\pm 50,\label{eq:kns}
\end{eqnarray}
The last coefficient above is only an
 estimate~\cite{LeDiberder:1992fr} but it is often used to evaluate
 the uncertainty resulting from the missing higher order terms.

Finally, the resulting non-universal $\beta$-function coefficients in
 the $\alpha_\tau$-scheme are then given by:
\begin{eqnarray}
  \beta_{\tau,2} & = & \beta_2^{\overline{\rm MS}}{\nf3} +
  \left(-\frac{299}{6} + 36\,\zeta_3\right)\beta_1{\nf3}
+ \left(\frac{169}{144}-\frac{1}{3}\pi^2\right)
    \beta_0^3{\nf3}\nonumber\\ & & +
    \left(\frac{26713}{36}+472\,\zeta_3+600\,\zeta_5
    -1296\,\zeta^2_3\right)\beta_0{\nf3}\nonumber\\ & = &
    \frac{79813}{16} + 6552\,\zeta_3 + 5400\,\zeta_5 - 243\,\pi^2 -
    11664\,\zeta^2_3 \simeq -789, \nonumber\\ \beta_{\tau,3} & = &
    \beta_3^{\overline{\rm MS}}{\nf3}
    +\left(-\frac{299}{3}+72\,\zeta_3\right) \beta_2^{\overline{\rm
    MS}}{\nf3} +
    \left(\frac{89401}{36}-3588\,\zeta_3+1296\,\zeta_3^2\right)\beta_1\nonumber\\
    & & + \frac{1819}{432}\beta_0^4{\nf3}
+ \left(\frac{845}{144}
-\frac{5}{3}\pi^2\right)\beta_1{\nf3}\,\beta_0^2{\nf3}
+ \left(-\frac{12673115}{27} + 555556\,\zeta_3 \right.\nonumber\\ & &
    \left.\phantom{\frac{1}{2}} - 179400\,\zeta_5 +
    129600\,\zeta_3\,\zeta_5 + 101952\,\zeta_3^2 
 - 186624\,\zeta_3^3
    + 128\,K_4^{\overline{\rm MS}}\right) \beta_0{\nf3}\nonumber\\ & =
& -\frac{585179735}{144} + 4820288\,\zeta_3 - 1614600\,\zeta_5 +
1166400\,\zeta_3\,\zeta_5 \nonumber\\ & & + 1000512\,\zeta^2_3 -
8640\,\pi^2 - 1679616\,\zeta^3_3 + 1152\, K_4^{\overline{\rm MS}}
\simeq-46776+1152\, K_4^{\overline{\rm MS}}.
\label{eq:beta_tau}
\end{eqnarray}
Our results agree with the numerical results published
in~\cite{Korner:2000xk}.


\begin{thebibliography}{99}

\bibitem{Shifman:bx}
M.~A.~Shifman, A.~I.~Vainshtein and V.~I.~Zakharov,
Nucl.\ Phys. {\bf B147}, 385 (1979).

\bibitem{Beneke:1998ui}
M.~Beneke,
Phys.\ Rept. {\bf 317}, 1 (1999)
[arXiv:hep-ph/9807443].

\bibitem{Beneke:1994qe}
M.~Beneke and V.~M.~Braun,
Phys.\ Lett. {\bf B348}, 513 (1995)
[arXiv:hep-ph/9411229].

\bibitem{Ball:1995ni}
P.~Ball, M.~Beneke and V.~M.~Braun,
Nucl.\ Phys. {\bf B452}, 563 (1995)
[arXiv:hep-ph/9502300].

\bibitem{Dokshitzer:1995qm}
Y.~L.~Dokshitzer, G.~Marchesini and B.~R.~Webber,
Nucl.\ Phys. {\bf B469}, 93 (1996)
[arXiv:hep-ph/9512336].

\bibitem{Grunberg:1980ja}
G.~Grunberg,
Phys.\ Lett. {\bf B95}, 70 (1980)
[Erratum-ibid. {\bf B110}, 501 (1982)].

\bibitem{Grunberg:1982fw}
G.~Grunberg,
Phys.\ Rev. {\bf D29}, 2315 (1984).

\bibitem{Brodsky:1998mf}
S.~J.~Brodsky, M.~S.~Gill, M.~Melles and J.~Rathsman,
Phys.\ Rev. {\bf D58}, 116006 (1998)
[arXiv:hep-ph/9801330].

\bibitem{Brodsky:1999fr}
S.~J.~Brodsky, M.~Melles and J.~Rathsman,
Phys.\ Rev. {\bf D60}, 096006 (1999)
[arXiv:hep-ph/9906324].

\bibitem{Brodsky:1994eh}
S.~J.~Brodsky and H.~J.~Lu,
Phys.\ Rev. {\bf D51}, 3652 (1995)
[arXiv:hep-ph/9405218].

\bibitem{Broadhurst:1993ru}
D.~J.~Broadhurst and A.~L.~Kataev,
Phys.\ Lett. {\bf B315}, 179 (1993)
[arXiv:hep-ph/9308274].

\bibitem{Brodsky:1995tb}
S.~J.~Brodsky, G.~T.~Gabadadze, A.~L.~Kataev and H.~J.~Lu,
Phys.\ Lett. {\bf B372}, 133 (1996) [arXiv:hep-ph/9512367].

\bibitem{Brodsky:2000cr}
S.~J.~Brodsky, E.~Gardi, G.~Grunberg and J.~Rathsman,
Phys.\ Rev. {\bf D63}, 094017 (2001)
[arXiv:hep-ph/0002065].

\bibitem{Rathsman:2001xe}
J.~Rathsman,
in {\it Proc. of the 5th International Symposium on Radiative Corrections (RADCOR 2000) } ed. Howard E. Haber, arXiv:hep-ph/0101248.

\bibitem{Gardi:2001wg}
E.~Gardi and G.~Grunberg,
Phys.\ Lett. {\bf B517}, 215 (2001)
[arXiv:hep-ph/0107300].

\bibitem{Bjorken_Drell} J.D. Bjorken and S.D. Drell,
{\it Relativistic Quantum Fields}, McGraw-Hill, New-York, 1965.

\bibitem{Brodsky:1982gc}
S.~J.~Brodsky, G.~P.~Lepage and P.~B.~Mackenzie,
Phys.\ Rev. {\bf D28}, 228 (1983).

\bibitem{Hornbostel:2002af}
K.~Hornbostel, G.~P.~Lepage and C.~Morningstar,
hep-ph/0208224 (2002).

\bibitem{Brodsky:1997jk}
S.~J.~Brodsky and P.~Huet,
Phys.\ Lett. {\bf B417}, 145 (1998)
[arXiv:hep-ph/9707543].

\bibitem{Susskind:pi}
L.~Susskind,
in {\it Les Houches 1976, Proceedings, Weak and Electromagnetic Interactions At High Energies}, 207-308, Amsterdam 1977.

\bibitem{Appelquist:es}
T.~Appelquist, M.~Dine and I.~J.~Muzinich,
Phys.\ Rev. {\bf D17}, 2074 (1978).

\bibitem{pinch1}
J.~M.~Cornwall,
Phys.\ Rev. {\bf D26}, 1453 (1982).

\bibitem{pinch2}
J.~M.~Cornwall and J.~Papavassiliou,
Phys.\ Rev. {\bf D40}, 3474 (1989).

\bibitem{Watson:1996fg}
N.~J.~Watson,
Nucl.\ Phys. {\bf B494}, 388 (1997)
[arXiv:hep-ph/9606381].

\bibitem{Binosi:2002vk}
D.~Binosi and J.~Papavassiliou,
hep-ph/0209016 (2002).

\bibitem{Braaten:hc}
E.~Braaten,
Phys.\ Rev.\ Lett. {\bf 60}, 1606 (1988).

\bibitem{Braaten:1988ea}
E.~Braaten,
Phys.\ Rev. {\bf D39}, 1458 (1989).

\bibitem{Narison:1988ni}
S.~Narison and A.~Pich,
Phys.\ Lett. {\bf B211}, 183 (1988).

\bibitem{Braaten:1991qm}
E.~Braaten, S.~Narison and A.~Pich,
Nucl.\ Phys. {\bf B373}, 581 (1992).

\bibitem{LeDiberder:1992fr}
F.~Le Diberder and A.~Pich,
Phys.\ Lett. {\bf B289}, 165 (1992).

\bibitem{Neubert:1995gd}
M.~Neubert,
Nucl.\ Phys. {\bf B463}, 511 (1996)
[arXiv:hep-ph/9509432].

\bibitem{Maxwell:1996ig}
C.~J.~Maxwell and D.~G.~Tonge,
Nucl.\ Phys. {\bf B481}, 681 (1996)
[arXiv:hep-ph/9606392].

\bibitem{Girone:1995xb}
M.~Girone and M.~Neubert,
Phys.\ Rev.\ Lett. {\bf 76}, 3061 (1996)
[arXiv:hep-ph/9511392].

\bibitem{Barate:1998uf}
(ALEPH Collaboration), R. Barate et al.,
Eur.\ Phys.\ J. {\bf C4}, 409 (1998).

\bibitem{OPAL}
(OPAL~Collaboration), K. Ackerstaff et al.,
\EUJ {C7}, 571 (1999)

\bibitem{Korner:2000xk}
J.~G.~K{\"o}rner, F.~Krajewski and A.~A.~Pivovarov,
Phys.\ Rev. {\bf D63}, 036001 (2001)
[arXiv:hep-ph/0002166].

\bibitem{Milton:2000fi}
K.~A.~Milton, I.~L.~Solovtsov, O.~P.~Solovtsova and V.~I.~Yasnov,
Eur.\ Phys.\ J. {\bf C14}, 495 (2000)
[arXiv:hep-ph/0003030].

\bibitem{Cvetic:2001sn}
G.~Cvetic and T.~Lee,
Phys.\ Rev. {\bf D64}, 014030 (2001)
[arXiv:hep-ph/0101297].

\bibitem{Cvetic:2001ws}
G.~Cvetic, C.~Dib, T.~Lee and I.~Schmidt,
Phys.\ Rev. {\bf D64}, 093016 (2001)
[arXiv:hep-ph/0106024].

\bibitem{Marciano:vm}
W.~J.~Marciano and A.~Sirlin,
Phys.\ Rev.\ Lett.  {\bf 61}, 1815 (1988).

\bibitem{Barnett:1996hr}
(Particle Data Group Collaboration), R.~M.~Barnett et al.,
Phys.\ Rev. {\bf D54}, 1 (1996).

\bibitem{Brodsky:1998ua}
S.~J.~Brodsky, J.~R.~Pel{\'a}ez and N.~Toumbas,
Phys.\ Rev. {\bf D60}, 037501 (1999)
[arXiv:hep-ph/9810424].

\bibitem{Menke}
S.~Menke,
Nucl.~Phys.~(Proc.~Suppl.) {\bf B76}, 299 (1999),
and {\bf B86}, 196 (2000).

\bibitem{Ioffe:2000ns}
B.~L.~Ioffe and K.~N.~Zyablyuk,
Nucl.\ Phys. {\bf A687}, 437 (2001)
[arXiv:hep-ph/0010089].

\bibitem{Geshkenbein:2001mn}
B.~V.~Geshkenbein, B.~L.~Ioffe and K.~N.~Zyablyuk,
Phys.\ Rev. {\bf D64}, 093009 (2001)
[arXiv:hep-ph/0104048].

\bibitem{PDG}
(Particle Data Group Collaboration), K.~Hagiwara et al.,
Phys.\ Rev. {\bf D66}, 010001 (2002).

\bibitem{Mattingly:ej}
A.~C.~Mattingly and P.~M.~Stevenson,
Phys.\ Rev. {\bf D49}, 437 (1994)
[arXiv:hep-ph/9307266].

\bibitem{Hoyer:2000ca}
P.~Hoyer and J.~Rathsman,
JHEP {\bf 0105}, 020 (2001)
[arXiv:hep-ph/0011209].

\bibitem{Howe:2002rb}
D.~M.~Howe and C.~J.~Maxwell,
Phys.\ Lett. {\bf B541}, 129 (2002)
[arXiv:hep-ph/0204036].

\bibitem{Shirkov:2001sm}
D.~V.~Shirkov,
Eur.\ Phys.\ J. {\bf C22}, 331 (2001)
[arXiv:hep-ph/0107282].

\bibitem{Brodsky:1989pv}
S.~J.~Brodsky and G.~P.~Lepage,
{\bf SLAC-PUB-4947} (1989).

\bibitem{Brodsky:1997dh}
S.~J.~Brodsky, C.~R.~Ji, A.~Pang and D.~G.~Robertson,
Phys.\ Rev. {\bf D57}, 245 (1998)
[arXiv:hep-ph/9705221].

\bibitem{Lepage:1980fj}
G.~P.~Lepage and S.~J.~Brodsky,
Phys.\ Rev. {\bf D22}, 2157 (1980).

\bibitem{Gross:1973id}
D.~J.~Gross and F.~Wilczek,
Phys.\ Rev.\ Lett. {\bf 30}, 1343 (1973).

\bibitem{Politzer:fx}
H.~D.~Politzer,
Phys.\ Rev.\ Lett. {\bf 30}, 1346 (1973).

\bibitem{Caswell:1974gg}
W.E.~Caswell,
Phys.\ Rev.\ Lett. {\bf 33}, 244 (1974).

\bibitem{Jones:mm}
D.~R.~Jones,
Nucl.\ Phys. {\bf B75}, 531 (1974).

\bibitem{Egorian:1978zx}
E.~Egorian and O.~V.~Tarasov,
Theor.\ Math.\ Phys. {\bf 41}, 863 (1979)
[Teor.\ Mat.\ Fiz. {\bf 41}, 26 (1979)].

\bibitem{Tarasov:au}
O.~V.~Tarasov, A.~A.~Vladimirov and A.~Y.~Zharkov,
Phys.\ Lett. {\bf B93}, 429 (1980).

\bibitem{Larin:tp}
S.~A.~Larin and J.~A.~Vermaseren,
Phys.\ Lett. {\bf B303}, 334 (1993)
[arXiv:hep-ph/9302208].

\bibitem{vanRitbergen:1997va}
T.~van Ritbergen, J.~A.~Vermaseren and S.~A.~Larin,
Phys.\ Lett. {\bf B400}, 379 (1997)
[arXiv:hep-ph/9701390].

\bibitem{Chetyrkin:bj}
K.~G.~Chetyrkin, A.~L.~Kataev and F.~V.~Tkachov,
Phys.\ Lett. {\bf B85}, 277 (1979).

\bibitem{Dine:1979qh}
M.~Dine and J.~R.~Sapirstein,
Phys.\ Rev.\ Lett. {\bf 43}, 668 (1979).

\bibitem{Celmaster:xr}
W.~Celmaster and R.~J.~Gonsalves,
Phys.\ Rev.\ Lett. {\bf 44}, 560 (1980).

\bibitem{Gorishnii:1990vf}
S.~G.~Gorishnii, A.~L.~Kataev and S.~A.~Larin,
Phys.\ Lett. {\bf B259}, 144 (1991).

\bibitem{Surguladze:1990tg}
L.~R.~Surguladze and M.~A.~Samuel,
Phys.\ Rev.\ Lett. {\bf 66}, 560 (1991)
[Erratum-ibid. {\bf 66}, 2416 (1991)].

\end{thebibliography}
\end{document}